# Compression Stress Effect on Dislocations Movement and Crack propagation in Cubic Crystal

Suprijadi[a], Ely Aprilia[a], Meiqorry Yusfi[b]

[a]Faculty of Mathematics and Natural Sciences, Institut Teknologi Bandung, Jl. Ganesha 10, Bandung 40132 Indonesia. Email: supri@fi.itb.ac.id, elyaprilia@yahoo.co.id
[b]Department of Physics, Andalas University, Kampus Limau Manis, Padang 25163, Indonesia.
Email: meiqorry@yahoo.com

**Abstract.** Fracture material is seriously problem in daily life, and it has connection with mechanical properties itself. The mechanical properties is belief depend on dislocation movement and crack propagation in the crystal. Information about this is very important to characterize the material. In FCC crystal structure the competition between crack propagation and dislocation wake is very interesting, in a ductile material like copper (Cu) dislocation can be seen in room temperature, but in a brittle material like Si only cracks can be seen observed. Different techniques were applied to material to study the mechanical properties, in this study we did compression test in one direction. Combination of simulation and experimental on cubic material are reported in this paper. We found that the deflection of crack direction in Si caused by vacancy of lattice, while compression stress on Cu cause the atoms displacement in one direction. Some evidence of dislocation wake in Si crystal under compression stress at high temperature will reported.

## 1 Introduction

Fracture material is seriously problem in daily life, in the past and present the accident by a crack propagation growth rapidly cause of material ages. Many problem in the past was initiate by a small crack which is growth and propagate to whole body, for example in airplane crash Boeing 747-146SR which flight from Tokyo (Haneda) to Osaka (Itami) on 1985, the aircraft was crash at Mt. Osutaka by fatigue metal and crack propagation only less from 1 hour after take-off from Haneda[1], sink of 12 US Ships including SS *John P. Gaines* which sank on 24 November 1943 [2] and other big damage as in dam break [3]. Base on this problems, many investigation with different techniques were done to study the phenomena. In principle, almost of investigation technique using stress or shear, for example stress were use in three point bending [4], hardness test [5] and thin plate buckling experiment [6].

Knowledge of material properties is useful in determining how the treatment is allowed in order not to damage the material. Loading which may result in damage to the materials is granted if the excessive load applied to them. If a material is given the excessive load will result in changes the properties of the materials. The ability of material in receiving excessive load that resulted in changes in shape or deformation makes the material can be classified into two parts, brittle and ductile [7]. In ductile material, dislocation wake is very easy to observe in room temperature, for example in steel [8] founded that dislocations more observed in unannealed steel crack propagation mechanism in Al thin film is reported [9], while in brittle material only crack were observed [10]. Increasing temperature on brittle material caused a dislocation wake, and a dislocation is believed can be occurred from a crack tip [11].

To increase the understanding of material properties, a computation on failure analysis were done by many researchers using molecular dynamic methods [12], or other methods [13]. In this paper we did combination of computation and experiment to understanding those phenomena, the propagation of cracks and dislocation wake will be reported. Possibility of crack propagation and its competition with dislocation wake will be discussed.

## 2. Experimental

Experiments on real materials were done to see the crystal damage in cubic crystal. A 500μm x 4mm x 15mm thin plate of cubic crystal was pressed under continuous force with speed less than 1mm/minute as schematic in Fig.1. The sample (in this figure the sample in Copper (Cu)) will bend in the fixture as can be seen in Fig.1.b, Different material were used, Si with (111) plane is used as represent of brittle material, while Cu used as represent of ductile material. To observe the crystal deformation on surface, combination of etch pitch and optical microscope were applied.

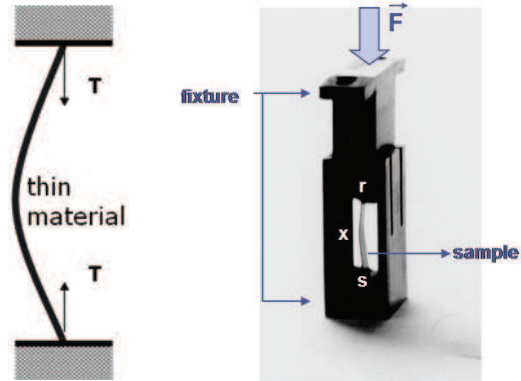

(a) Schematic    (b) sample in special fixture
Figure 1.Experiment apparatus

## 3. Simulation

In generally, the crack front will propagate if any atom loss their cohesion force, the atomic decohesion can be happen if there is an energy which bigger than critical energy. Solving for the critical energy release rate of atomic decohesion can be done completely by assuming an infinitesimal distribution of opening displacement running along the crack front develops under applied loading. The applied stress solutions are now the vertical tensile components of the stress solutions. The opening stress vs. opening displacement relationship is given by [13]

$$\sigma[\Delta(r)] = \frac{2\gamma_S}{L^2} \Delta(r) e^{\frac{-\Delta(r)}{L}}$$

where $\gamma_s$ is the surface energy and L is the opening displacement at the peak of the stress displacement relationship. This formula must also be expressed in terms of δ, the displacement occurring at the plane of the crack front in order to solve the integral equilibrium equation expression. Varying the crack length and crack tip root radius demonstrates the effect of the pre-crack geometry parameters on the critical energy release rate of atomic decohesion.

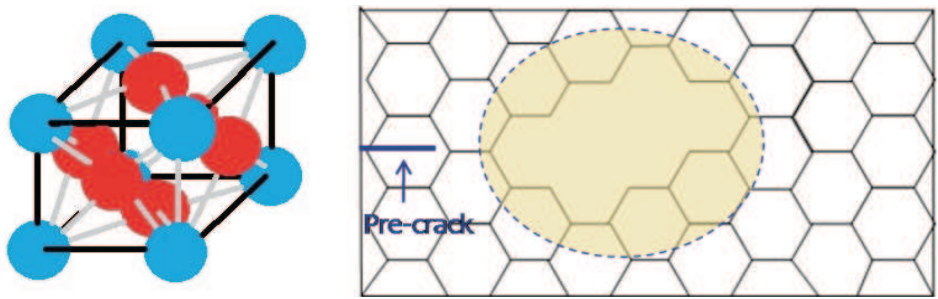

(a)   Face Center Cubic (FCC) of Cu    (b) (111) Si
*Figure 2, FCC lattice and crystal plane model*

Different methods for study of material damage were applied. To study of dislocation in cubic metal the molecular dynamic based on Wei [12] algorithm was used in cooper (Cu) with FCC structure (Fig.1.a), the results was analyzed using mathematical tools freeware Scilab 5.2.2 (www.scilab.org). While for study crack propagation in brittle material was simulated using finite element in two dimensions Si with orientation in [111], the planes model can be seen in Fig.1.b. The [111] geometry choose because of simple structure and has the same length between closer atoms.

Simulation on crack follows the Griffith criteria, that crack will propagate if stress in a direction higher than stress maximum, the algorithm for the crack simulation as follows:

```
Start subprogram
Initiation ;define pre-crack, F_external, MaterialLength
calculate VectorPosition, CriticalForce
While (F_external > CriticalForce)
  discritization
  while (VectorPosition < MaterialLength)
        find DOF ;degree of freedom
        find stress distribution each node
        set new VectorPosition
  end while
end while
end subprogram
```

In simulation of crack propagation in brittle material, the Si will be use (Fig.2.b). Silicon has Young's modulus (E)=18.5 x $10^{11}$ dyne/$cm^2$ , Surface energy($\gamma$) is 1000 erg/$cm^2$; Poison ratio ($\nu$) is 0.26, and 5.43 Å of crystal dimension.

## 4. Results and discussion

*4.1 Brittle material*
Figure 3 shows the crack front of Si under stressed. In Fig.a, the experiment was done in temperature lower than transition temperature. We observed that only crack propagated along stress axis while increasing temperature just above the temperature transition [11], the dislocation start growth up, in Fig.3.b, shows TEM image of dislocation wake around the crack and crack tip as reported in [17].

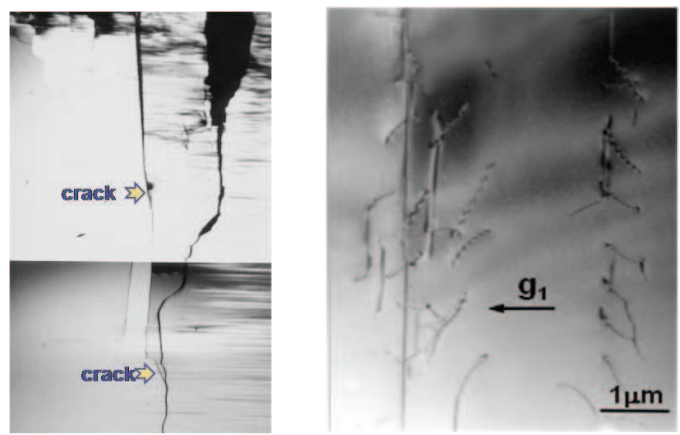

(a) cracks          (b) dislocation wake
Figure 3 Plastic deformations in Si under compression stress

Simulation results that, the crack will propagate along stress direction, because of the (111) plane, we found that the crack propagated in zigzag pattern (Fig.4). To control the crack direction, at beginning, we setup the crack tip (Fig.2.b), different length of crack tip shows the crack easier deviated. Figure 4.a, shows the zigzagged crack by crack tip, deviated cracks more bigger in 100nm pre-crack (denote by (ii)) compare to 50nm denote by (i). All of the simulations did in perfect crystal, no vacancy as shown in shadow area in Fig.2.b.

To find out effect of imperfect crystals, vacancy in some of node did in the middle of structure (Fig.2.b). The results show that increasing number of vacancy will make the crack easier to deviate. As reference, perfect crystal crack propagation (Fig. 4.a.) is used and in Fig.4.b, denoted by (i). Two atoms (ii) and four atoms (iii) was take out to give vacancy in the middle of structure (Fig.2.b). Increasing numbers of atom will bigger deflected as shown in Fig.4.b. The same problems were found in previous works as reported [7] as crack-step, the deflected crack were found in interface of $Si_3N_4$ and BN[14], the crack deflected cause of increasing energy during propagation in interface layer.

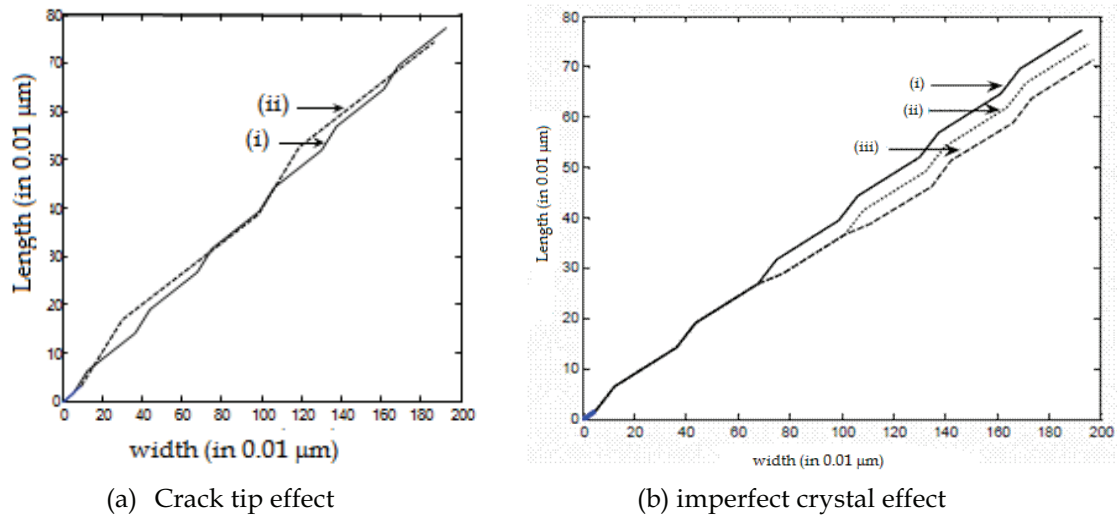

(a) Crack tip effect　　　　　　　　　　　　(b) imperfect crystal effect

Figure 4. Crack propagation in brittle material

*4.2 Ductile material*
Experiment on ductile material, we use the Cu, the sample was stressed in compression machine using continue stress with speed of 2 mm/minutes. The stress is stopped and release just after the sample buckle, it can be seen in stress-strain curve as below (Fig.5.a). In the center of buckling sample, we observed using optical microscope. Figure 5.b shows the image of dislocations in several places (see circles and denote by d1, d2, and d3), the dislocation occur in the same direction with stress direction (show by arrow). It is shown that the copper start deformation in room temperature under $1.938 \times 10^8$ N/m² (194 MPa).

Figure 6, shows the result of molecular dynamic simulation in Cu under constant stress, different stress was applied start from 100 MPa to 4000 MPa. The figure shows the atoms configuration before pressed (a) and after pressed (b) the atoms were collected in the middle of cube, but it is not to easy to analyze, the atom displacement can be plot as Fig.4, it can be seen at higher stress, the atoms displacement increase rapidly. Atoms were distributed in one axis follow the compressing stress, the figure shows the dislocations start in area closed to surface and move to inner of cube. This phenomenon was reported in simulation of dislocation wake in bulk crystal [15].

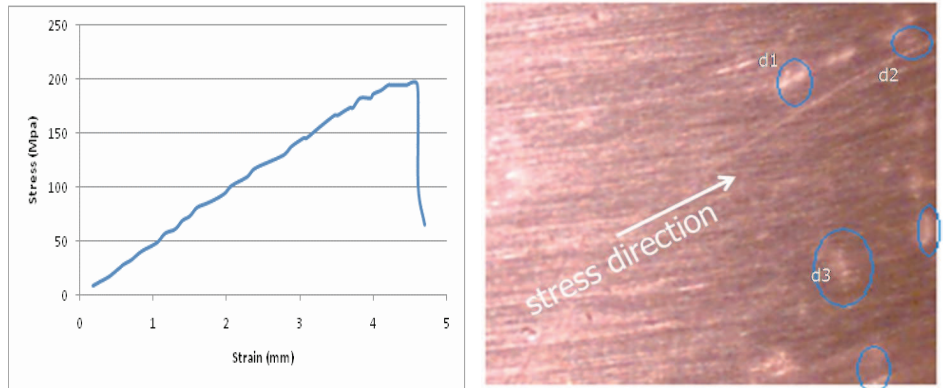

(a)   Stress-strain curve          (b) optical microscope in Cu
Figure 5 Experimental results of compressed copper

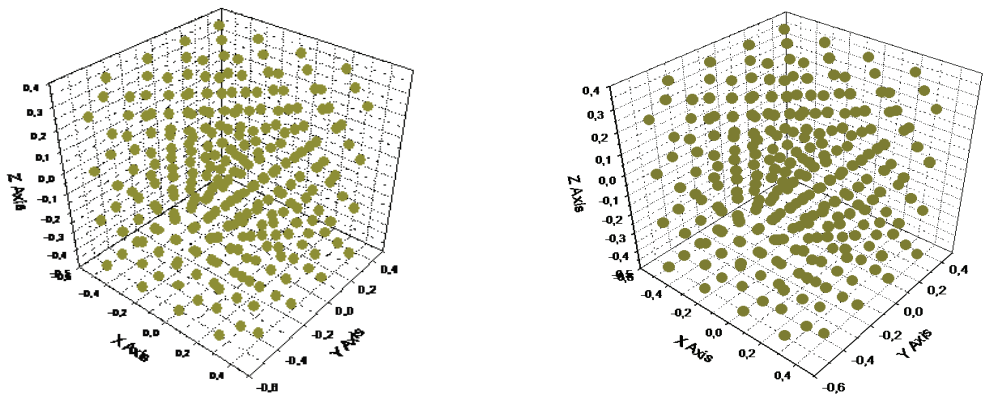

(a)      Before stressed          (b) After stressed with 4000 MPa along z-axis
Figure 6 Atoms distribution in cube metal

The movement of the dislocation is made of a succession of motion of each straight segment, and the kinetics of motion of a straight segment depends upon the local effective stress and the line tension. The local effective stress is a sum of threshold stress, internal stress, external stress tensor, and image force tensor [15]. In this model, effect of external stress tensor is set to be main factor. Figure 5, shows the dislocation movement to central of the sample, it explained that the simulation give the same results as founded in other simulations method [15]. Figure 7.a, shows that if the stress applied not big enough, the dislocation only occur in small area close to surface where the stress concentrated. Different phenomenon were reported, the dislocation nucleated close to crack tip in Si at ductile brittle transition temperature (DBTT) [8] or in Cu [16], if there is a crack, the dislocation will starting occur just close to the tip, and spread in majority as stress direction applied.

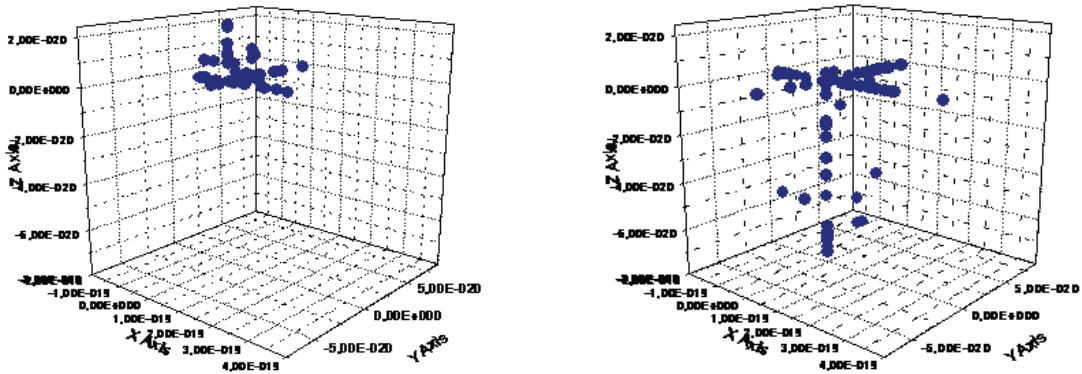

(a) Stressed by 1000 MPa　　　　　　　　　(b) Stressed by 4000 MPa
Figure 5 Distributed of atoms displacement in Cu

5. Conclusion

In brittle material, some vacancy atoms in a cubic crystal make the crack propagate in deviate direction along the compression direction. It can be shown that number of vacancies has tight relation with deviate angle of crack propagation, while the pre-crack play important role on crack propagation. In the ductile material, the dislocation wake starting from just beneath surfaces of material to middle of sample as found in [15], while the effect of compression direction has tight relation with dislocation wake.